\documentclass[aps,pre,showpacs,twocolumn,amsmath,amssymb]{revtex4-1}

\usepackage{graphicx}
\usepackage{dcolumn}
\usepackage{bm}
\usepackage{amsmath}
\usepackage{amssymb}
\usepackage{latexsym}
\usepackage{epsfig}
\usepackage{amsbsy}
\usepackage{array}
\usepackage{amssymb}
\usepackage{setspace}
\usepackage{bm}
\begin{document}

\title{Long-lived quantum speedup based on plasmonic hot spot systems}

\author{Jun Ren$^{1,2}$}

\author{Tian Chen$^{1*}$\footnote{Email:xbwang@mail.tsinghua.edu.cn}}

\author{Xiangdong Zhang$^1$}

\affiliation{$^{1}$Beijing Key Laboratory of Nanophotonics ${\&}$ Ultrafine Optoelectronic Systems, School of Physics, 
Beijing Institute of Technology, 100081, Beijing, China\\
$^2$College of Physical Science and Information Engineering, Hebei Normal University, 050024, Shijiazhuang, Hebei, China\\
}
\begin{abstract}
Long-lived quantum speedup serves as a fundamental component for quantum algorithms. The quantum walk is identified as an ideal scheme to realize the long-lived quantum speedup. However, one finds that the duration of quantum speedup is very short in real systems implementing quantum walk. The speedup can last only dozens of femtoseconds in the photosynthetic light-harvesting system, which was regarded as the best candidate for quantum information processing. Here, we construct one plasmonic system with two-level molecules embodied in the hot spots of one-dimensional nanoparticle chains to realize the long-lived quantum speedup. The coherent and incoherent coupling parameters in the system are obtained by means of Green's tensor technique. Our results reveal that the duration of quantum speedup in our scheme can exceed 500 fs under strong coherent coupling conditions, which is several times larger than that in the photosynthetic light-harvesting system. Our proposal presents a competitive scheme to realize the long-lived quantum speedup, which is very beneficial for quantum algorithms.
\end{abstract}


\maketitle
\section{Introduction}
Quantum information exhibits the advantage over its classical counterpart due to the appearance of the quantum speedup [1-12]. Long-lived quantum speedup has been widely applied in the quantum information processing, e.g., quantum algorithms. One ideal theoretic scheme involving the long-lived quantum speedup is the quantum walk [13, 14]. The mean squared displacements of excitation in the ideal one-dimensional (1D) quantum walks display the ballistic spreading $\langle (\Delta x)^2\rangle \propto t^2$, Such rate of spreading in the quantum walk is indicated as the ideal quantum speedup, which was commonly attributed to the quantum coherence in the systems [13-15]. In comparison, the corresponding classical random walk characterizes the diffusive spreading $\langle (\Delta x)^2\rangle \propto t$ and does not possess the quantum speedup.

In recent years, many experiments have demonstrated the existence of quantum coherence in natural systems. For example, the ultrahigh efficient transport due to the quantum coherence has been observed in photosynthetic light-harvesting systems [16-23]. These systems are treated as potential platforms to realize the continuous-time quantum walk and quantum speedup algorithms. Nevertheless, the long-lived quantum coherence in photosynthetic light-harvesting systems cannot ensure the long-lived quantum speedup. Hoyer and coauthors [24] pointed out that, due to the disorder and dephasing, the transition from ballistic to diffusive spreading occurs at about 70 fs in photosynthetic light-harvesting complexes even though the quantum coherence lasts much longer. The short duration of quantum speedup hinders photosynthetic light-harvesting systems to be employed as a platform in the realization of quantum speedup algorithms.

Recently, the quantum coherence has also been found in nanophotonic systems such as nanocavities [25, 26], photonic crystals [27] and plasmonic systems [28]. As addressed in Refs.~[28-31], the coupling resonances among nanoparticle trimer could induce the strong couplings between the molecules, and the strong quantum coherence has been revealed between two molecules in a symmetrical nanoparticle trimer system [28]. Given that the quantum coherence is the key factor for constructing the quantum walks, these works motive us to study how to realize the ideal quantum walks based on strong coupling plasmonic nanostructures and explore the quantum speedup in these nanostructures.

In this work, we demonstrate that 1D continuous-time quantum walk can be constructed within a scheme of the nanoparticle chain involving plasmonic hot spots. The decoherence from the ambient environment has been taken into account. The dynamics for continuous-time quantum walk in such a system is obtained by means of the Lindblad master equation approach and the electromagnetic Green's tensor technique. Our results reveal that, due to the strong nearest-neighbor coupling between the molecules, the duration of quantum speedup in our proposed system can reach 500 fs, which is several times larger than the duration of quantum speedup in the photosynthetic light-harvesting system. Our implementation of continuous-time quantum walk based on such plasmonic nanostructures presents a new platform to realize the quantum speedup algorithms.

The rest of this paper is arranged as follows: in Sec.~II, we present the general description for the plasmonic hot spot system. The correspondence between the plasmonic hot spot system and the continuous-time quantum walk has been provided. Then in Sec.~III, we provide the dynamics of the plasmonic hot spot system with different frequencies. The study of quantum speedup in our plasmonic hot spot system is addressed in Sec.~IV. Further discussions with different parameters of the chain, the number of molecules and the nonlocal effect are presented in Sec.~V. Finally, we make a summary in Sec.~VI.

\section{GENERAL DESCRIPTIONS IN 1D PLASMONIC NANOPARTICLE CHAIN}

The 1D nanoparticle chain is depicted in Fig.~1, in which the separation distances between the particles are assumed as the same (denoted by $d$). The six two-level molecules (marked by $1\sim6$) are inserted into the gaps of the chain, and the orientations of the electric dipole moments of these molecules are assumed along the axis of the chain. Such configuration is also indicated as the plasmonic hot spot system [28]. Here, we consider one excitation in these two-level molecules. The six molecules can be mapped to the positions in the 1D quantum walk (see Fig.~1). Though we focus on the dynamics of the molecular system which is composed by six molecules in the main text, we also study the system with other numbers of molecules and provide the discussion in Supplemental Material. Here, the position of the $i$th molecule is expressed by $i$, and the displacement $x_i$ of the ith molecule is given by its position in the 1D chain which is indicated in Fig.~1. The mean-squared displacement of the excitation in these molecules can be obtained as
\begin{equation}
\langle (\Delta x)^2\rangle=\mathrm{Tr}(\rho(\Delta x)^2),
\end{equation}
where the $\Delta x$ denotes the displacement between the current and initial position of the molecules. The density matrix $\rho$ describes the density matrix of all molecules. In the following, we will study the evolution of these molecules and provide the correspondence between the dynamics of the excitation in the molecules and 1D quantum walk.

The inevitable dissipation from the metal nanoparticles and radiation into the free space should be considered for studying the dynamics of the molecules. The spreading speed of excitation and distance relate closely to the dynamics of the molecules. The dynamics of the six molecules can be described in the form of the Lindblad master equation [20, 32]


\begin{figure}[b]
\includegraphics{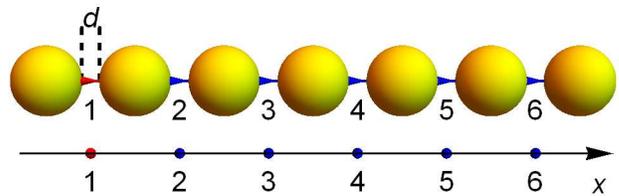}
\caption{\label{fig:epsart} The 1D nanoparticle chain with six two-level molecules inserted in the gaps. The number 1 to 6 represents the $1$st to $6$th two-level molecule. Initially the excitation locates in the first molecule.}
\end{figure}
\begin{figure}[b]
\includegraphics{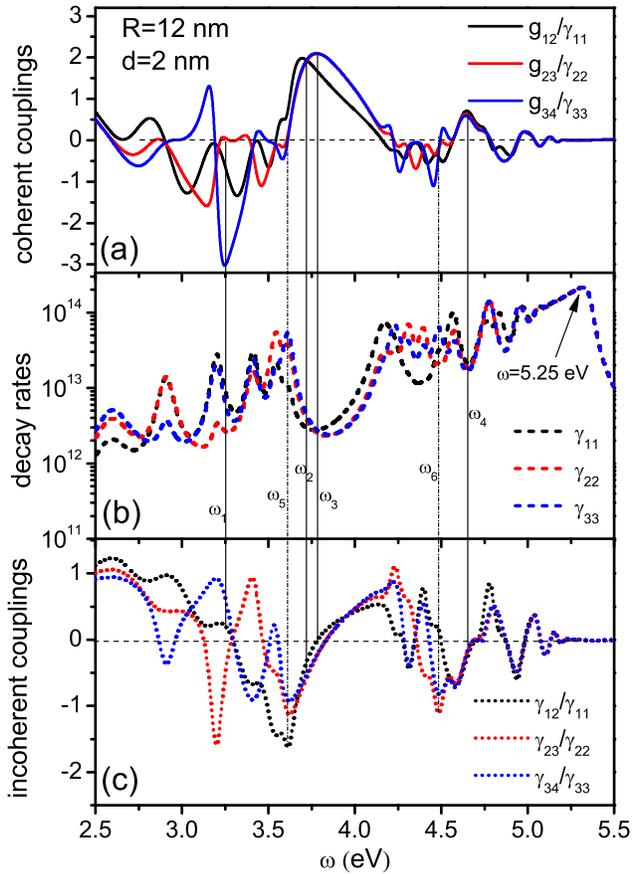}
\caption{\label{fig:epsart} (a) The nearest-neighbor coherent coupling strengths $g_{12}/\gamma_{11}$, $g_{23}/\gamma_{22}$ and $g_{34}/\gamma_{33}$. (b) Decay rates of single molecule. (c) The nearest-neighbor incoherent coupling strengths. The parameters of the system is $R=12$ nm, $d=2$ nm, the dielectric function of Ag spheres is taken as Drude model ($\omega_p=9.01$ eV, $\gamma=0.05$ eV).}
\end{figure}
\begin{figure*}
\includegraphics{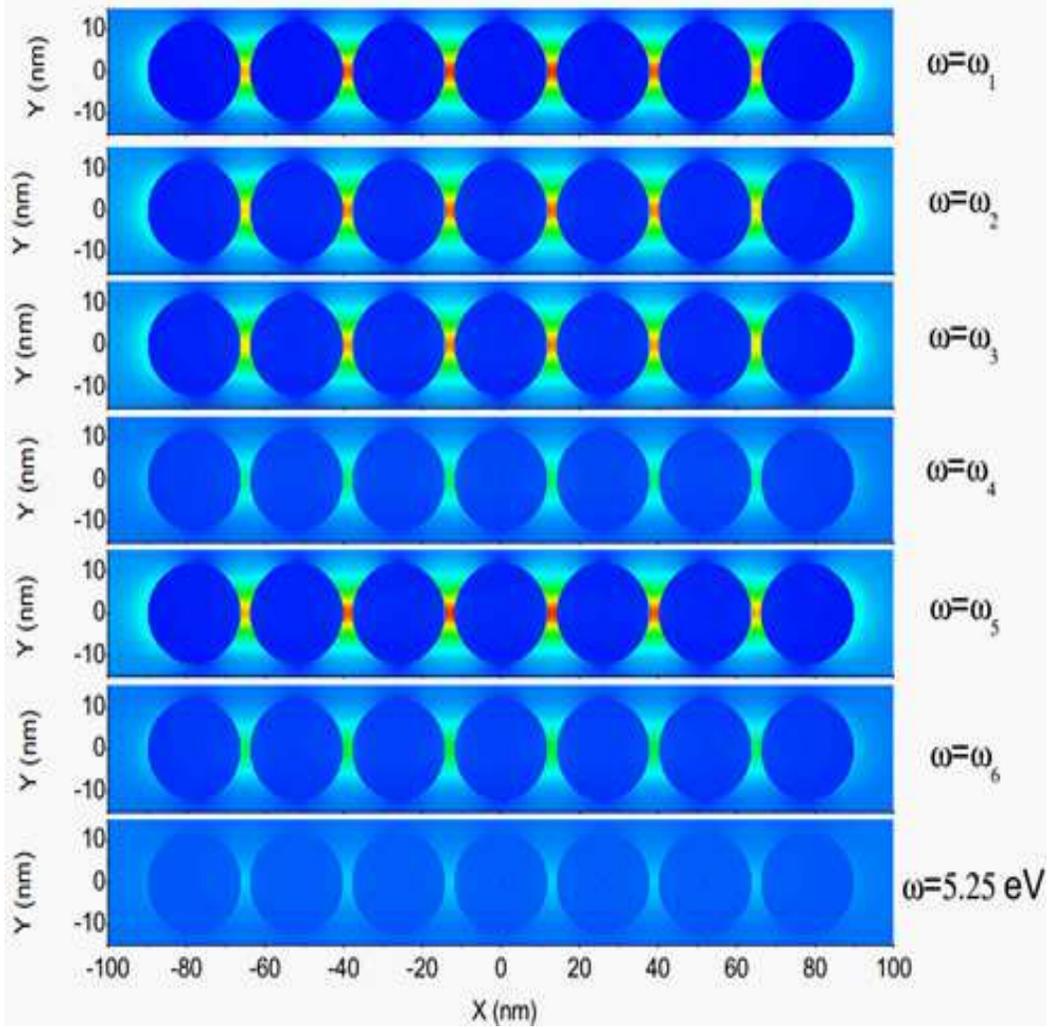}
\caption{\label{fig:wide} The electric field amplitude pattern in the 1D nanoparticle chain with different frequencies for the normal incidence. The parameters of the system are the same as that in Fig. 2.}
\end{figure*}
\begin{figure}[b]
\includegraphics{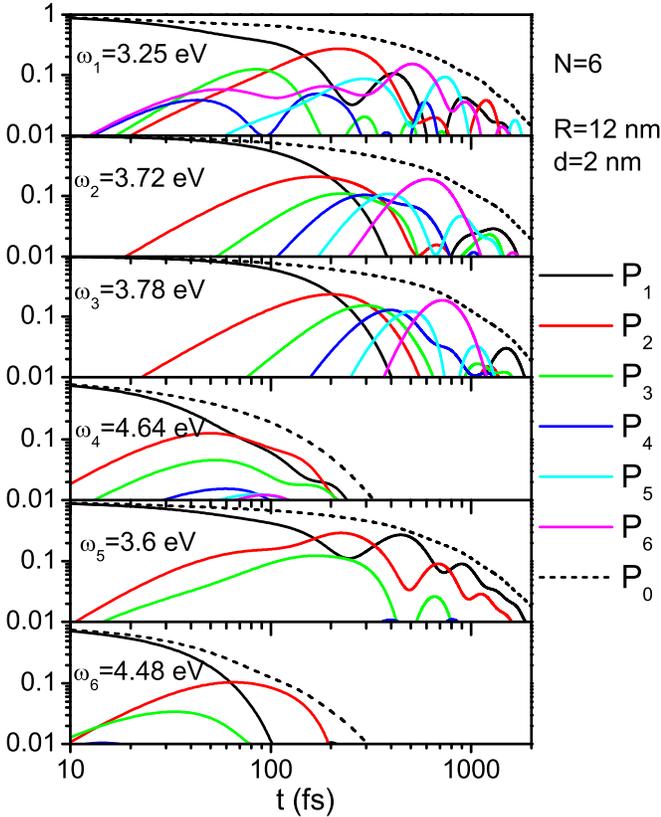}
\caption{\label{fig:epsart} The log-log plot of the time evolution of the populations of six molecules when the first molecule is excited and other five molecules are in their ground states initially. Where the solid lines $P_1\sim P_6$ denote the populations of the six molecules, and the dashed black line $P_0$ is the time evolution of population residing in all molecules. The parameters of the system are the same as Fig.~2.}
\end{figure}
\begin{figure}[b]
\includegraphics{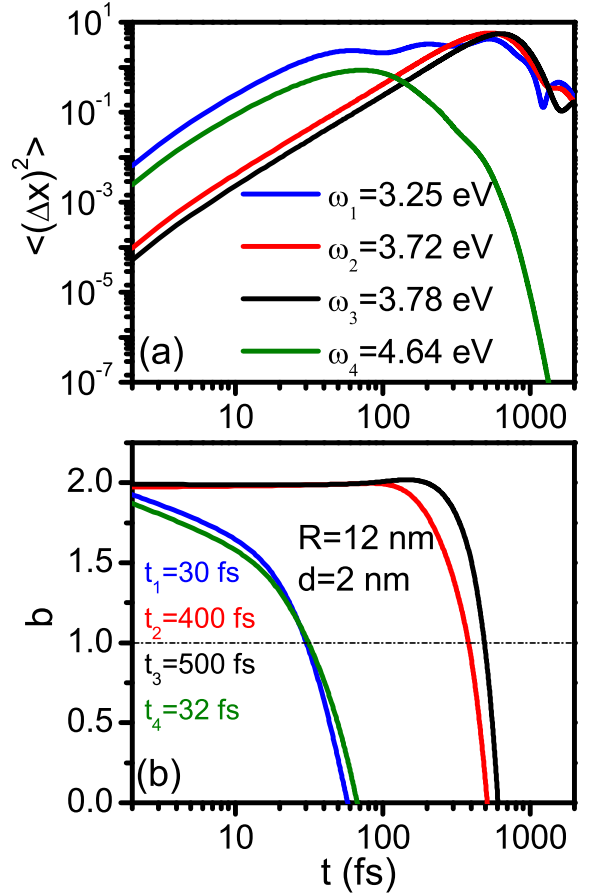}
\caption{\label{fig:epsart} (a) The mean squared displacement of excitation $\langle (\Delta x)^2\rangle$ and (b) the fitted power $b$ as a function of time. The cases are presented with blue ($\omega=\omega_1$), red ($\omega=\omega_2$), black ($\omega=\omega_3$) and green ($\omega=\omega_4$) lines respectively. Other parameters are the same as Fig. 2.}
\end{figure}
\begin{figure*}
\includegraphics{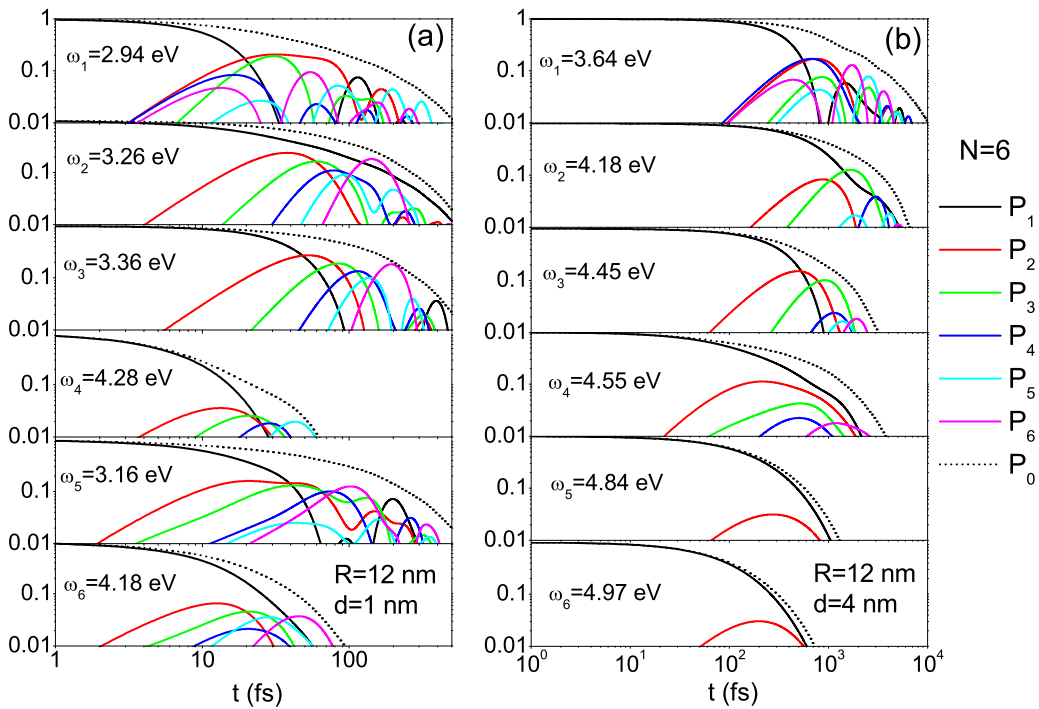}
\caption{\label{fig:wide}The log-log plot of the time evolution of the populations of six molecules when the first molecule is excited and other five molecules are in their ground states initially. Where the solid lines $P_1\sim P_6$ denote the populations of the six molecules, and the dashed black line $P_0$ is the time evolution of population residing in all molecules. The separation distance $d=1$ nm (a) and $d=4$ nm (b). The radii of the spheres is fixed at $R=12$ nm.}
\end{figure*}
\begin{figure}[b]
\includegraphics{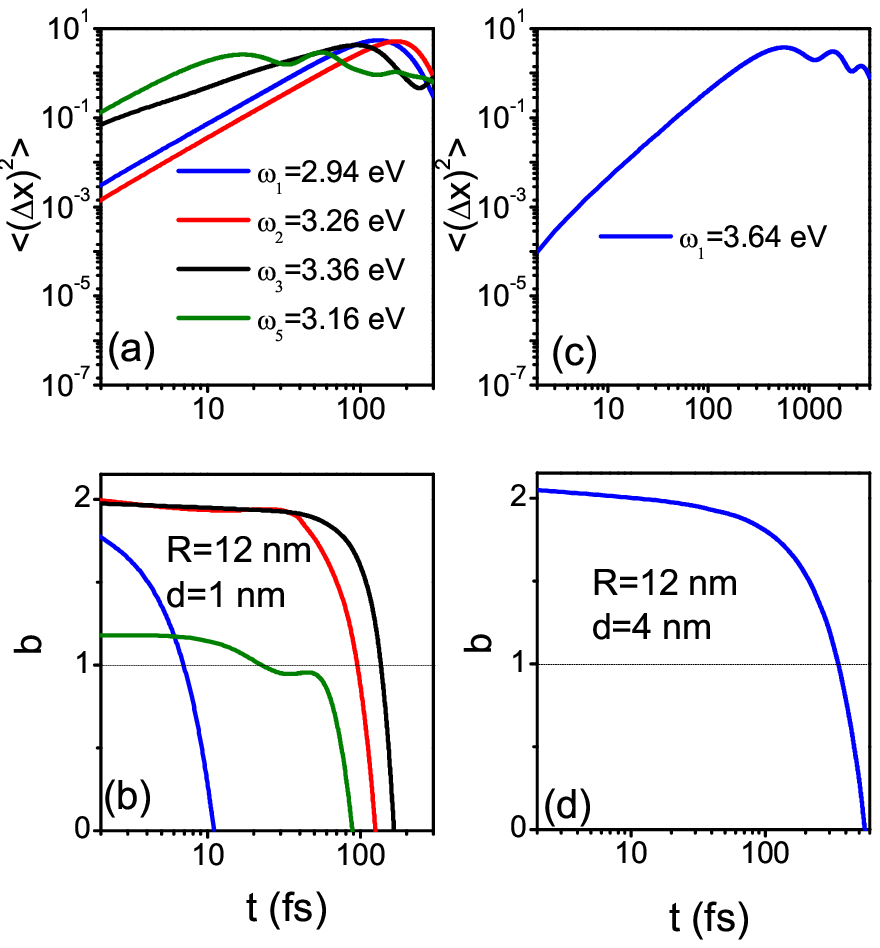}
\caption{\label{fig:epsart} The mean squared displacement of excitation $\langle (\Delta x)^2\rangle$ [(a) and (c)] and the fitted power $b$ [(b) and (d)] as a function of time when $d=1$ nm (left column) and $d=4$ nm (right column). The cases are presented with blue ($\omega=\omega_1$), red ($\omega=\omega_2$), black ($\omega=\omega_3$) and green ($\omega=\omega_4$) lines respectively.}
\end{figure}

\begin{equation}
\begin{split}
\partial_t\rho=&\frac{i}{\hbar}[\rho,H]\\
&+\sum_{i=1}^N\frac{\gamma_{ii}}{2}(2\sigma_i\rho\sigma_i^\dag-\sigma_i^\dag\sigma_i\rho-\rho\sigma_i^\dag\sigma_i)
\\&+\sum_{i\not=j}^N\frac{\gamma_{ij}}{2}(2\sigma_i\rho\sigma_j^\dag-\sigma_i^\dag\sigma_j\rho-\rho\sigma_i^\dag\sigma_j),
\end{split}
\end{equation}
where $\sigma_i^\dag$ and $\sigma_i$ are the creation and annihilation operators of the $i$th molecule, $\gamma_{ij}$ represents the interference term (incoherent coupling) between the $i$th and the $j$th molecules, which could lead to the decay of the off-diagonal elements of density matrix $\rho$. The $\gamma_{ii}$ represents the dissipation of the $i$th molecule, which includes the dissipation in the environment and non-radiation transition. In the metal nanoparticle cluster environment, the dissipation arises mainly from the loss in metal. The Hamiltonian of these molecules is described as
\begin{equation}
H=\sum_{i=1}^N\hbar\omega_0\sigma_i^\dag\sigma_i+\sum_{j>1}^Ng_{ij}(\sigma_i^\dag\sigma_j+\sigma_j^\dag\sigma_i),
\end{equation}
Here, $N$ is the number of molecules, and $N=6$ is taken in this work. The parameter $\omega_0$ is the transition frequency of molecule, and here for simplicity, we assume they have the same transition frequency. The coefficient $g_{ij}$ in Hamiltonian is the coherent coupling strength between the $i$th and the $j$th molecules,
\begin{equation}
g_{ij}=\frac{1}{\pi\varepsilon_0\hbar}\mathcal{P}\int_0^\infty\frac{\omega^2 \mathrm{Im}[\vec{\mu}_i^*\cdot\tensor{G}(\vec{r}_i,\vec{r}_j,\omega)\cdot\vec{\mu}_j]}{c^2(\omega-\omega_0)}\mathrm{d}\omega,
\end{equation}
where $\mathcal{P}$ stands for the principle integral, and $\varepsilon_0$ is the relative dielectric constant of the background medium, in this work the background is taken as water ($\varepsilon_0=1.77$). Here $\tensor{G}$ is the total Green's tensor. To avoid the sophisticated principle integral, the $g_{ij}$ can be simplified as [28, 33]
\begin{equation}
g_{ij}=\frac{\omega_0^2}{\varepsilon_0\hbar c^2}\mathrm{Re}[\vec{\mu}_i^*\cdot\tensor{G}^s(\vec{r}_i,\vec{r}_j,\omega)\cdot\vec{\mu}_j],
\end{equation}
where $\tensor{G}^s$ represents the scattered Green's tensor of the nanoparticle chain, and the detailed calculations can be found in Appendix. With the electromagnetic Green's tensor technique, the incoherent coupling strength $\gamma_{ij}$ appearing Eq.~(2) can be obtained as
\begin{equation}
\gamma_{ij}=\frac{2\omega_0^2}{\varepsilon_0\hbar c^2}\mathrm{Im}[\vec{\mu}_i^*\cdot\tensor{G}(\vec{r}_i,\vec{r}_j,\omega)\cdot\vec{\mu}_j],
\end{equation}
in which the calculation of the total Green's tensor $\tensor{G}$ is addressed in Appendix. According to Eq.~(5), the dissipation in the $i$th molecule can be expressed as $\gamma_{ii}=2\omega_0^2/(\varepsilon_0\hbar c^2)\mathrm{Im}[\vec{\mu}_i^*\cdot\tensor{G}(\vec{r}_i,\vec{r}_i,\omega)\cdot\vec{\mu}_i]$. The calculation procedure for $\gamma_{ii}$ can be found in Appendix.

In this work, all the interactions among the six molecules have been included, from the nearest neighbor interactions between the $1$st and $2$nd molecules, to the interaction between the $1$st and $6$th molecules. According to the symmetry of the system, the interactions (coherent and incoherent) between the $1$st and $2$nd molecules are the same to the interactions between the $5$th and $6$th molecules, namely $g_{12}=g_{56}$ and $\gamma_{12}=\gamma_{56}$, and for the other nearest neighbor interactions, there is $g_{23}=g_{45}$ and $\gamma_{23}=\gamma_{45}$. Similarly, for the non-nearest-neighbor interactions we have $g_{13}=g_{46}$ and $\gamma_{13}=\gamma_{46}$, $g_{24}=g_{35}$ and $\gamma_{24}=\gamma_{35}$, $g_{14}=g_{36}$ and $\gamma_{14}=\gamma_{36}$, $g_{15}=g_{26}$ and $\gamma_{15}=\gamma_{26}$. The detailed calculation method for $g_{ij}$ and $\gamma_{ij}$ can be found in Appendix.

In the following, we will investigate the correspondence between the dynamics of molecules in hot spot system and the standard quantum walk. By employing the quantum trajectory method introduced in Refs.~[20, 21], we can re-express the dynamics of the excitation in molecules as
\begin{equation}
\partial_t\rho=\frac{i}{\hbar}(\rho^\dag H_\mathrm{eff}^\dag-H_\mathrm{eff}\rho)+\sum_{i\not=j}^N\gamma_{ij}\sigma_i\rho\sigma_j^\dag+\sum_{i=1}^N\gamma_{ii}\sigma_i\rho\sigma_i^\dag,
\end{equation}
where the effective Hamiltonian can be written as
\begin{equation}
H_\mathrm{eff}=H-\frac{i}{2}\sum_{i\not=j}^N\gamma_{ij}\sigma_i^\dag\sigma_j-\frac{i}{2}\sum_{i=1}^N\gamma_{ii}\sigma_i^\dag\sigma_i.
\end{equation}
Notice that $H_\mathrm{eff}$ is a non-Hermitian Hamiltonian, and it commutes with excitation number operator $\mathcal{N}=\sum_{i=1}^{N}\sigma_i^\dag\sigma_i$, thus it preserves the number of excitation, and can only give rise to the jump of excitation between different molecules. The second term in the right of Eq.~(7) originates from the incoherent interaction between the two-level molecules and plasmons, it can induce dephasing in these molecules without changing the number of excitations. These two terms mentioned above induce the jumps in the single-excitation manifold and no jump to the other excitation manifolds. The last term in the right of Eq.~(7) originates from the dissipation in the plasmonic environment and radiation to the free space, it can change the number of excitations and generate jumps between excitation manifolds. When neglecting the jumps to other excitation manifolds, we have the master equation in the case of a no jump trajectory as
\begin{equation}
\partial_t\rho_\mathrm{eff}=\frac{i}{\hbar}(\rho_\mathrm{eff}^\dag H_\mathrm{eff}^\dag-H_\mathrm{eff}\rho_\mathrm{eff})+\sum_{i\not=j}^N\gamma_{ij}\sigma_i\rho_\mathrm{eff}\sigma_j^\dag,
\end{equation}
where $\rho_\mathrm{eff}$ is the effective density operator describing the density matrix of all molecules. This equation is often considered as a directed quantum walk on the single-excitation manifold described by the density operator $\rho_\mathrm{eff}$ [21, 34]. In our discussion below, we will focus on the dynamics of density operator $\rho$ including the jumps between excitation manifolds [i.e., Eq.~(7)]. We will show that even when influenced by the ambient environment, our system can still possess long-lived quantum speedup.

\section{THE QUANTUM WALK DYNAMICS IN 1D PLASMONIC HOT SPOT SYSTEMS}
According to the theory in Sec.~II, the dynamics of our system involving one excitation depends closely on the coherent and incoherent coupling strengths among the molecules. In our study, the nanoparticles are taken as the Ag spheres. For the dielectric functions of Ag, the Drude model is adopted ($\omega_p=9.01$ eV and $\gamma=0.05$ eV). In Fig.~2, when the radii of the Ag spheres are taken as $R=12$ nm, and the separation distances between two nearest-neighbor spheres are fixed at $d=2$ nm , we present the coherent and incoherent coupling strengths and molecule decay rates as a function of the transition frequency of molecule. Such configuration could be realized by using small molecules that link the particles in a line, like the experimental research shown in Ref.~[30], in which the distance can be as small as about $1$ nm. In Fig.~2(a), the nearest-neighbor coherent couplings $g_{12}/\gamma_{11}$, $g_{23}/\gamma_{22}$ and $g_{34}/\gamma_{33}$ are showed by black, red and blue lines, respectively. The three lines in Fig.~2(b) correspond to the decay rates of the first, second and third molecules ($\gamma_{11}$, $\gamma_{22}$ and $\gamma_{33}$). As for Fig.~2(c), the three lines correspond to the nearest-neighbor incoherent couplings $\gamma_{12}/\gamma_{11}$, $\gamma_{23}/\gamma_{22}$ and $\gamma_{34}/\gamma_{33}$, respectively.

It is seen clearly that, there exist multiple resonances in three panels of Fig.~2. These resonances can affect the dynanics of the excitation of molecules. To clarify the properties of the resonances, we focus on some special cases that correspond to different coupling conditions. These special frequencies are marked by $\omega_1\sim\omega_6$  in Fig.~2, in which $\omega_1\sim\omega_4$ are the cases that the system possesses the large coherent couplings and small incoherent couplings. For example, when $\omega=3.25$ eV (marked by $\omega_{1}$), one of the nearest-neighbor coherent couplings ($g_{12}/\gamma_{11}$) can reach to $-3$, however the other two are small. And when $\omega=3.72$ eV and 3.78 eV (marked by $\omega_{2}$ and $\omega_{3}$), the nearest-neighbor coherent couplings ($g_{12}/\gamma_{11}$, $g_{23}/\gamma_{22}$ and $g_{34}/\gamma_{33}$) can reach to 2 simultaneously. When $\omega=4.64$ eV (marked by $\omega_4$), the three coherent couplings are 0.8 simultaneously. In contrast, $\omega_5$ and $\omega_6$ are the cases that coherent couplings are very small and incoherent couplingss are at resonance peaks.

To further clarify the properties of the resonances marked in Fig.~2, in Fig.~3 we plot the corresponding electric field patterns of the nanoparticle system for the normal incidence, in which $\omega_1\sim\omega_6$ correspond to the six frequencies marked in Fig.~2. It is shown that when the couplings (coherent or incoherent) are in maximums ($\omega=\omega_1\sim\omega_3$ and $\omega_5$), the electric field patterns are similar, and the fields in hot spots are much larger than other region. This stems from the coupling resonances among the nanoparticle chain as has been discussed in Refs.~[28, 29]. In comparison, we also plot the field pattern under the single scattering frequency ($\omega=5.25$ eV as marked in Fig.~2(b)), which is showed in the last panel of Fig. 3. In this panel, the intensities of the fields in the gaps are about the same with that around the particles. That means the fields in the molecules cannot be confined. When $\omega=\omega_4$ and $\omega_6$, due to the weak coupling resonances in the system, the fields in hot spots are weaker compared to the cases of $\omega_1\sim\omega_3$ and $\omega_5$ due to the weak coupling resonances in the system, which are similar to the case of single scattering resonance.

The electric field patterns in Fig.~3 show that strong resonance couplings only exist in the coupling resonance region between the nanoparticles, which confines the large electric field within the positions of hot spots. Such strong confinement of fields results in the strong couplings between the molecules that located at the hot spots. In addition, comparing Fig.~2(c) with Fig.~2(b) we find that, incoherent couplings between the molecules correspond very well to the resonance peaks, which are resonant couplings; while coherent couplings have a shift compared to resonance peaks, which are off-resonant couplings. [28]

Various resonance couplings can lead to different populations of the excitation of molecules. The  time evolution of the populations of six molecules (denoted by $P_1\sim P_6$) are shown in Fig.~4 with six different frequencies (marked by $\omega_1\sim\omega_6$ as shown in Figs.~2 and 3). It has to be pointed out that, in the calculations of populations (Fig.~4), all the non-nearest-neighbor couplings among molecules have also been included. The results of the non-nearest-neighbor coherent and incoherent coupling strengths between the molecules have been shown in Supplementary Material. In Fig.~4, the dashed black line denoted by P0 is the time evolution of population in all molecules. As addressed in Fig.~4, when $\omega=3.25$ eV ($\omega_1$), the excitation can reach every molecule even though  $P_4$ and $P_5$ are very small. At this frequency, only one nearest-neighbor coupling $g_{34}/\gamma_{33}$ can reach to $-3$, the other nearest-neighbor couplings $g_{12}/\gamma_{11}$ and $g_{23}/\gamma_{22}$ are very small. The imbalance of the coherent couplings leads to small populations at the $4$th and $5$th molecules. Even though, the time of the energy residing in molecules is larger than 1000 fs in this case (dashed black line in the first panel). In comparison, when the transition frequency is taken as $\omega=3.72$ eV ($\omega_2$) or 3.78 eV ($\omega_3$), all the nearest-neighbor coherent couplings showed in Fig.~2 can reach 2 simultaneously, and the nearest-neighbor incoherent couplings and decay rates are small enough. In these cases, the molecules inserted in the gap of the nanoparticle cluster can be treated as an ideal 1D chain with same nearest-neighbor couplings. In Fig.~4, the populations on all molecules at the transition frequency $\omega=3.72$ eV ($\omega_2$) or 3.78 eV ($\omega_3$). Thus under these frequencies, all molecules in the 1D chain can be excited, and $P_0$ is similar to the first panel.

When the transition frequency is taken as $\omega=4.64$ eV ($\omega_4$), the coherent and incoherent couplings are all small (see Fig.~2). In this case the excitation cannot transport (see the fourth panel in Fig.~4) any more, and the percentage of the energy residing in molecules decreases to zero quickly, thus the quantum walk cannot be constructed. When $\omega=3.6$ eV ($\omega_5$), according to the Figs.~2,~4 and 5, we find that the nearest-neighbor and non-nearest-neighbor incoherent couplings are in their peak values, and nearest-neighbor coherent couplings are almost zero. In this case, although the population in all molecules lasts a long time similar to the first three cases, the excitation can only reach to the third molecule and cannot transport to the other three molecules. When $\omega=4.48$ eV ($\omega_6$), the nearest-neighbor incoherent couplings are also in their peak values but much smaller than the case of $\omega_5$, and the nearest-neighbor coherent couplings are also very small. In this case, the excitation almost cannot transport any more, and the energy residing in the molecules decreases to zero quickly. Thus, when the transition frequency is taken as $\omega=\omega_4$, $\omega_5$ and $\omega_6$, due to very small nearest-neighbor coherent couplings, not all molecules inserted in the gap of nanoparticles can be excited. In these three cases, an ideal 1D chain with same nearest-neighbor couplings cannot be constructed, and the dynamics of the excitations are far different from the dynamics of quantum walk. 

\section{THE QUANTUM SPEEDUP IN PLASMONIC HOT SPOT SYSTEMS}
In this section, we explore under what conditions our hot spot system holds the quantum speedup as in the standard quantum walk. A natural way to evaluate the quantum speedup is the exponent $b$, which relates to the spreading of the excitation in the 1D chain by the power law $\langle (\Delta x)^2\rangle \propto t^b$. Here, $b=2$ corresponds to the ideal quantum speedup in the quantum walk, $b=1$ corresponds to the diffusive transport of classical random walks and $b<1$ corresponds to the sub-diffusive transport [24]. The long-lived quantum speedup is one of the most important property in the ideal quantum walk, which makes it exhibit advantages over the classical walk in information processing. In one realistic system, the duration of $b>1$ determines the quality of the quantum speedup in quantum walks. In Fig.~5(a), we present the mean squared displacement of excitation $\langle (\Delta x)^2\rangle$ as a function of time, in which the parameters of the system are the same to those in Fig.~2. In Fig.~5(b), we plot the the fitted power $b$ as a function of time at four frequencies. In Fig.~5(a) and (b), the frequencies $\omega_1\sim\omega_4$ correspond to the cases that nearest-neighbor coherent couplings are in their peak values.

According to Figs.~2 and 4, when the frequency is taken as $\omega=\omega_2$ or $\omega_3$, the nearest-neighbor coherent couplings between any two adjacent molecules in 1D chain are strong, and the other couplings can be neglected. In these two cases, these molecules along 1D chain have the same structure as that in the continuous-time quantum walk, and populations at all molecules can reach the nearly same value in sequence. Our calculations in Fig.~6(b) shows that the duration of quantum speedup under $\omega_2$ and $\omega_3$ can reach to about $400\sim500$ fs, which is several times larger than those in photosynthetic light-harvesting systems (about 70 fs) [24]. At these frequencies, the molecules inserted in the gap of 1D nanoparticles form a realistic 1D quantum walk system with long-lived quantum speedup.

In the following, we explore how the separation distance between adjacent nanoparticles affects the excitation in the systems. When the separation distance changes, the competition between couplings and dissipations in the system could largely affect the duration of quantum speedup. We investigate the two cases that $d=1$ nm and $d=4$ nm, respectively. In both cases, the radii of nanoparticles is same, $R=12$ nm.

The solid lines in Fig.~6(a) and (b) show that the time evolution of the populations of six molecules in different frequencies for $d=1$ nm and $4$ nm, respectively. And the dashed black lines in Fig. 6 denote the time evolution of population in all molecules. It is clear that the energy dissipates much more quickly when d=1 nm than the case that $d=4$ nm due to the much larger dissipation. The calculations of nearest-neighbor coupling strengths and non-nearest-neighbor coupling strengths are addressed in Supplementary Material. Similar to Fig.~4 (the populations for $d=2$ nm), the frequencies $\omega_1\sim\omega_4$ in Fig.~6 correspond to the cases that nearest-neighbor coherent coupling strengths are in their peak values and incoherent couplings are almost zero. We first focus on the case with $d=1$ nm [Fig.~6(a)]. When the transition frequency is $\omega_2$ or $\omega_3$, nearest-neighbor coherent coupling strengths can reach 1.8 simultaneously, which are a little smaller than the case with $d=2$ nm [there the nearest-neighbor coherent coupling strengths reach 2.0 in Fig. 2(a)]. While, due to large dissipations for the case with $d=1$ nm, populations on all molecules are dissipated in a very short time. When the transition frequency is taken as $\omega_1$ or $\omega_4$, coherent coupling strengths between the adjacent molecules are much smaller than the cases of $\omega_2$ and $\omega_3$. When the frequency is $\omega_5$ or $\omega_6$, all nearest-neighbor incoherent coupling strengths between adjacent molecules are in their peak values and coherent coupling strengths between them are almost zero. Under those transition frequencies ($\omega_1$, $\omega_4$, $\omega_5$ or $\omega_6$), due to very small nearest-neighbor coherent coupling strengths, the populations on molecules which are far away from the initially excited molecule are almost zero. In those cases, the dynamics of excitation among molecules along 1D chain is far different from that of quantum walk.

When the separation distance is $d=4$ nm (the associated coupling strengths can also be found in Supplementary Material), nearest-neighbor coherent coupling strengths between adjacent molecules only reach about 1.1 simultaneously, that is much smaller than the nearest-neighbor coherent coupling strengths with $d=1$ nm and $d=2$ nm. Notice that in the case with $d=4$ nm, the dissipation strengths (indicated in Supplementary Material) are one order of magnitude smaller than those with $d=2$ nm. Compared with the case with $d=1$ nm or 2 nm, when $d=4$ nm, the smaller dissipations in the molecules make populations survive in the system with longer time [as shown in Fig.~6(b)]. However, when $d=4$ nm, due to small coupling strengths between adjacent molecules, the excitation is nearly localized at the initial molecule under most frequencies, and the quantum walk dynamics is unable to be implemented.

Comparing Fig.~6(a) with Fig.~4 (the case with $d=2$ nm), we find that the surviving time of populations is shorten from about 1000 fs to 300 fs when the separation distance is changed from $d=2$ nm to 1 nm. Figure~8 shows the mean squared displacement of excitation $\langle (\Delta x)^2\rangle$ (a) and the fitted power $b$ (b) as a function of time when $d=1$ nm (left column) and $d=4$ nm (right column). Where the cases are presented with blue ($\omega=\omega_1$), red ($\omega=\omega_2$), black ($\omega=\omega_3$) and green ($\omega=\omega_4$) lines respectively. When the separation distance $d=1$ nm, we find that the duration of quantum speedup is no longer than 150 fs [see the case with the transition frequency chosen as $\omega_2$ or $\omega_3$ in Fig.~7(b)], which is much smaller than the duration of quantum speedup with $d=2$ nm [about 500 fs in Fig.~4(b)]. While when $d=4$ nm, the duration of quantum speedup can reach about 350 fs under $\omega=\omega_1$. Such duration of quantum speedup is comparable to that with $d=2$ nm. However, when the transition frequency satisfies $\omega=\omega_1$ with $d=4$ nm, we find that all molecules except the initially occupied molecule are nearly not excited [first panel of Fig.~6(b)]. Such small excitation at molecules cannot be seen as one efficient quantum walk.

\section{DISCUSSIONS}
In our study above, we have realized the long-live quantum speedup within the plasmonic hot spot system. The formation of quantum walk dynamics in this system can last hundreds of femtoseconds, which is one order magnitude larger than in the photosynthetic light-harvesting system. Based on discussions above, we find that among molecules inserted in our 1D nanoparticle chain, the competition between coherent/incoherent couplings and dissipations can largely influence the excitation of molecules. Such influence on the excitation leads to different quantum speedup in our system and affect the formation of the quantum walk within our scheme. From the view of quantum walk, the formation of standard quantum walk requires the same nearest-neighbor coupling strengths between any adjacent two sites. These same coupling strengths induce the walker to jump to the nearest sites with same probabilities and form the strong interference between the wave functions at every site. The probability in the central sites diminishes due to destructive interference and much more probability appears in the remote sites due to constructive interference, which generates the ideal quantum speedup in the standard quantum walk. When studying the quantum speedup in the plasmonic hot spot system, we find that the emergence of long-lived quantum speedup in the system indicates nearly same nearest-neighbor coupling strengths and very small non-nearest-neighbor coupling strengths between molecules [$\omega=3.72$ eV ($\omega_2$) or $3.78$ eV ($\omega_3$) in Fig.~4].

In the case with the radii of nanoparticle $R=12$ nm, the separation distance $d=2$ nm is an optimal solution for the ideal quantum walk with long-lived quantum speedup. For these parameters in the plasmonic hot spot system, we can obtain relatively same larger nearest-neighbor coupling strengths between any two adjacent molecules, and smaller non-nearest-neighbor coupling strengths between molecules. Similarly, when the radii of the nanoparticle is changed, the competition between the coherent/incoherent couplings and dissipations still exists in our scheme, and the optimal separation distance to construct quantum walk with long-lived quantum speedup could be changed. In Supplementary Material, we present coherent/incoherent coupling strengths and populations for the case with $R=10$ nm and $d=4$ nm. Comparing to the case with $R=12$ nm and $d=4$ nm, we find that the coherent coupling strengths decrease about 10\% when the radii of spheres decreases to $R=10$ nm. This directly leads to the decrease of populations on molecule. Therefore, when we decrease the radii of the spheres, we need to decrease the separation distance to ensure the large enough nearest-neighbor coherent couplings between molecules. On the contrary, when we increase the radii of the spheres in our scheme, dissipations in these molecules and coherent couplings between adjacent molecules increase simultaneously. In this case, the optimal separation distance also increases to implement a long-lived quantum speedup.

We also investigate the quantum speedup properties with different numbers of molecules. We fix the parameters of the hot spot system with the radii of sphere $R=12$ nm, and the distance between two adjacent spheres is $d=2$ nm. The numbers of molecules are chosen as $N=5$ and $N=7$, respectively. The calculated results have been presented in Supplementary Material. We find that the quantum walk properties in these systems are similar to the case with $N=6$, and simultaneously large nearest-neighbor coherent couplings lead to long-lived quantum speedup. In our discussion, we obtain that the optimal duration of quantum speedup is $420$ fs with the number of molecules $N=5$, and $580$ fs with $N=7$. Considering that the duration of quantum speedup is $500$ fs in our main text with the number of molecules $N=6$, for the plasmonic hot spot system with different numbers of molecules, we can realize the long-live quantum speedup.

In addition, when the radii of spheres decreases to 10 nm and separation distances are smaller than 1 nm, the nonlocal effect cannot be ignored [35-38]. In these cases, the plasmon-enhanced fluorescence could be decreased by a small value [39], namely dissipation strengths of molecules could reduce a little when the radii of spheres and separation distances between adjacent molecules are not too small. At the same time, previous researches have shown that coherent coupling strengths in a trimer system ($R=10$ nm and $d=1$ nm) only present small blue shifts due to the nonlocal effect [28], and values of coupling strengths keep the same. In our plasmonic hot spot system, we also investigate the nonlocal effect in Supplementary Material. We find that the nearest-neighbor couplings and dissipations only have blue shifts and have no obvious change in values. We can obtain the similar quantum speedup behavior with change of the molecular frequencies. It means that the nonlocal effect affects dissipations and coherent couplings among molecules little, and will not influence quantum speedup and the formation of quantum walk in our discussion.

\section{summary}
In this work, we investigate long-lived quantum speedup in molecules inserted in the gaps of 1D nanoparticle chain. Both the coherent and incoherent couplings among molecules have been included for the dynamics of the excitation of molecules, and the roles of nearest-neighbor and non-nearest-neighbor couplings on the quantum speedup have been analyzed. We have found that at some special frequencies, due to the large coupling resonance, the large nearest-neighbor coherent couplings between any two adjacent molecules and small dissipation of each molecule are obtained simultaneously. In this case, the dynamics of excitation among molecules in our scheme is similar to that of the continuous-time quantum walk, and the duration of quantum speedup in our scheme is several times larger than the duration in photosynthetic light-harvesting systems. Although in our study, the quantum speedup within 1D nanoparticle chain is addressed as the dynamics in 1D quantum walk on a chain, our scheme can also be extended to the case of quantum walk on a circle. Given that our discussion takes into the real experimental situation account (e.g., dissipation from the nanoparticles, radiation into free space, and so on), our proposal based on the plasmonic hot spot system presents a new scheme to have long-lived quantum speedup and provides a new platform to realize the continuous-time quantum walk under laboratory conditions.

\section*{Acknowledgement}
This work was supported by the National key R \& D Program of China under Grant No.~2017YFA0303800 and the National Natural Science Foundation of China (11574031 and 11604014). J. R. was also supported by the Hebei NSF under Grant No.~A2016205215.

\section*{appendix: The detailed calculations of coherent and incoherent coupling strengths among molecules in the nanoparticle (NP) chain}

Here we present the detailed calculation method of the coherent and incoherent terms ($g_{ij}$ and $\gamma_{ij}$) as shown in Eqs.~(4) and (5). In Eqs.~(4) and (5), $\tensor{G}^s(\vec{r}_i,\vec{r}_j,\omega)$ and $\tensor{G}(\vec{r}_i,\vec{r}_j,\omega)$ represent the scattered and total Green's tensors of the nanoparticle (NP) chain, respectively. The vectors $\vec{r}_i$ and $\vec{r}_j$ are the positions of the $i$th and $j$th molecules, respectively. The Green's tensor  $\tensor{G}(\vec{r}_i,\vec{r}_j,\omega)$ has the meaning of the total electric field in the position of $\vec{r}_i$ caused by a unit dipole at the location of $\vec{r}_j$ in the presence of NP chain, similar to $\tensor{G}^s(\vec{r}_i,\vec{r}_j,\omega)$. 

In this work, the Green's tensors of the NP system are all calculated with the multiple scattering T-matrix technique [40-42]. The total and scattered Green's tensors have the following relation
\begin{equation}
\tensor{G}(\vec{r}_i,\vec{r}_j,\omega)=\tensor{G}^s(\vec{r}_i,\vec{r}_j,\omega)+\tensor{G}^{vac}(\vec{r}_i,\vec{r}_j,\omega),
\end{equation}
where $\tensor{G}^{vac}(\vec{r}_i,\vec{r}_j,\omega)$ is the Green's tensor in the vacuum. In the following, we introduce the multiple scattering T-matrix method to calculate the Green's tensors mentioned above.

The incident field $\vec{E}_{inc}$ and the scattered field of the $i$th NP $\vec{E}_{s}^i$ can be expanded with the vector spherical functions (VSFs) [40-42]
\begin{equation}
\vec{E}_{inc}(\vec{r}-\vec{R}_i)=\sum_{\nu=1}^{\infty}a_\nu^i \vec{M}_\nu^1(k(\vec{r}-\vec{R}_i))
+b_\nu^i \vec{N}_\nu^1(k(\vec{r}-\vec{R}_i)),
\end{equation}
\begin{equation}
\vec{E}_{s}^i(\vec{r}-\vec{R}_i)=\sum_{\nu=1}^{\infty}f_\nu^i \vec{M}_\nu^3(k(\vec{r}-\vec{R}_i))
+g_\nu^i \vec{N}_\nu^3(k(\vec{r}-\vec{R}_i))\quad \left|\vec{r}_i\right|>\mathfrak{R}_i,
\end{equation}
where $\vec{M}_\nu^1$, $\vec{N}_\nu^1$, $\vec{M}_\nu^3$ and $\vec{N}_\nu^3$ are the well-known VSFs, and $\vec{r}_i$ is the position vector in the coordinate of the $i$th nanoparticle. $\mathfrak{R}_i$ is the radius of the smallest sphere circumscribing the $i$th object, in this work  $\mathfrak{R}_i$ is equal to the radii of the NPs.

The Green's tensor in the vacuum $\tensor{G}^{vac}(\vec{r}_i,\vec{r}_j,\omega)$ can be readily known in Ref.~[43]. In Eqs.~(A2) and (A3), the coefficients $a_\nu^i$, $b_\nu^i$, $f_\nu^i$ and  $g_\nu^i$ can be easily solved as soon as the form of the incident wave is given. The subscript  $\nu$ stands for ($m$, $n$) which are the indices of spherical harmonic functions. At the same time the internal field of the ith NP can be written as
\begin{equation}
\vec{E}_{int}^i(\vec{r}-\vec{R}_i)=\sum_{\nu=1}^{\infty}c_\nu^i \vec{M}_\nu^1(k(\vec{r}-\vec{R}_i))
+d_\nu^i \vec{N}_\nu^1(k(\vec{r}-\vec{R}_i)),
\end{equation}
According to the T-matrix method [47-49], $c_\nu^i$ and $d_\nu^i$ are related to $a_\nu^i$ and $b_\nu^i$ by the following matrix equation:
\begin{equation}\small
\begin{bmatrix}Q_i^{11}&Q_i^{12}\\
Q_i^{21}&Q_i^{22}\end{bmatrix} \begin{bmatrix}c^i\\d^i\end{bmatrix}\\
=\sum_{j=1}^{N(j\not=i)}
\begin{bmatrix}T_{ij}^{11}&T_{ij}^{12}\\
T_{ij}^{21}&T_{ij}^{22}\end{bmatrix} \begin{bmatrix}\mathrm{Rg}Q_j^{11}&\mathrm{Rg}Q_j^{12}\\
\mathrm{Rg}Q_j^{21}&\mathrm{Rg}Q_j^{22}\end{bmatrix} \begin{bmatrix}c^j\\d^j\end{bmatrix}+\begin{bmatrix}a^i\\b^i\end{bmatrix},
\end{equation}
where $Q_i^{pq}$ and $\mathrm{Rg}Q_j^{pq}$ are the T-matrix blocks for the $i$th and $j$th NPs, and  $T_{ij}^{pq}$ is the block of the transition matrix between the $i$th and the $j$th NPs [47-49]. By solving these equations, expansion coefficients of the inner field for each NP can be obtained. And also according to the matrix equation
\begin{equation}
\begin{bmatrix}f^i\\g^i\end{bmatrix}\\
=\begin{bmatrix}\mathrm{Rg}Q^{11}&\mathrm{Rg}Q^{12}\\
\mathrm{Rg}Q^{21}&\mathrm{Rg}Q^{22}\end{bmatrix} \begin{bmatrix}c^i\\d^i\end{bmatrix},
\end{equation}
the scattered expansion coefficients $f_\nu^i$ and $g_\nu^i$ of each NP can be easily calculated. The field outside the NPs then can be obtained using the following equation:
\begin{equation}
\vec{E}_{ext}=\vec{E}_{inc}+\sum_{i=1}^N \vec{E}_s^i,
\end{equation}
where $N$ is the number of nanoparticles.

In the following we calculate the external scattered field $\vec{E}_d^\kappa$ induced by a dipole source. We take the exciting source as a dipole $\vec{\mu}_{12}^\kappa$ located in $\vec{r}_\kappa$, the incident electric field can be expressed as
\begin{equation}
\vec{E}_{inc}=\frac{-\nabla\times\nabla\times\vec{A}_p^{(\kappa)}}{i\omega\mu_0\varepsilon_{med}\varepsilon_0},
\end{equation}
where
\begin{equation}
\vec{A}_p^{(\kappa)}=-i\omega\mu_0\frac{e^{ik\left|\vec{r}-\vec{r}_{\kappa}\right|}}{4\pi\left|\vec{r}-\vec{r}_{\kappa}\right|}\vec{\mu}_{12}^\kappa.
\end{equation}
Expanding Eq.~(A8) to the same form of Eq.~(A2), from Eqs.~(A3-A7) we can obtain the external scattered field $\vec{E}_{d}^{\kappa}=\sum_{i=1}^{N}\vec{E}_s^i$ caused by the dipole source. Then, the scattered Green's tensor $\tensor{G}^s(\vec{r}_i,\vec{r}_j,\omega)$ can be obtained with
\begin{equation}
\vec{n}_i\cdot\tensor{G}^s(\vec{r}_i,\vec{r}_j,\omega)\cdot\vec{n}_j=-\vec{n}_i\cdot\vec{E}_d^{\kappa}(\vec{r}_i)|_{\vec{r}_j}.
\end{equation}
where $\vec{n}_i$ and $\vec{n}_j$ are the unit vectors of the field and source dipole moments, respectively. $\vec{E}_d^{\kappa}(\vec{r}_i)|_{\vec{r}_j}$ represents the field in the position of $\vec{r}=\vec{r}_i$ induced by the source dipole that locates at $\vec{r}=\vec{r}_j$ in the presence of NP chain.

For the decay rate of the $i$th molecule $\gamma_{ii}$, the Green's tensor $\tensor{G}(\vec{r}_i,\vec{r}_j,\omega)$ represents the field generated by the dipole in its own position in the presence of NP chain, which can be calculated with
\begin{equation}
\vec{n}_i\cdot\tensor{G}(\vec{r}_i,\vec{r}_i,\omega)\cdot\vec{n}_i=-\vec{n}_i\cdot\vec{E}_d^{\kappa}(\vec{r}_i)|_{\vec{r}_i}.
\end{equation}
where $\vec{E}_d^{\kappa}(\vec{r}_i)|_{\vec{r}_i}$ represents the field induced by the source dipole that locates at   in its own position in the presence of NP chain.

{}

\end{document}